\documentclass[preprint,prb,amsmath]{revtex4-1}
\usepackage{color}
\usepackage{graphicx}

\begin{document}
\title{Effect of internal electric field on ferroelectric polarization in multiferroic TbMnO$_3$}
\author{Chandan De, Somnath Ghara and A. Sundaresan}
\email{sundaresan@jncasr.ac.in}
\address{Chemistry and Physics of Materials Unit and International Center for Materials Science, Jawaharlal Nehru Centre for Advanced Scientific Research, Jakkur P.O., Bangalore 560064 India}

\date{\today}

\begin{abstract}

Pyro-current measurements have been widely used to study ferroelectric properties in multiferroic materials. However, determination of intrinsic polarization by this method is not straightforward because of leakage current and trapped charge carriers. Here, we demonstrate the formation of internal electric field due to thermally stimulated charge carriers and its influence on ferroelectric polarization in a polycrystalline sample of the well known multiferroic TbMnO$_3$. While an electric field ($E_{ext}$) poling across the ferroelectric transition ($T_C$ $\sim$ 26 K) is essential to obtain depolarization current at $T_C$, the sample poled  only in the paraelectric state ($T_{pole}$ = 130 $-$ 50 K) also exhibits a pyro-current peak at $T_C$ but with the same polarity ($-$ $I_{pyro}$) as that of the external field ($-$ $E_{ext}$).  We demonstrate that these unusual behavior of pyro-current are caused by a positive internal electric field ($+$ $E_{int}$) which in turn is created by thermally stimulated free charge carriers during the poling process in the paraelectric state. We also show that a combination of DC-biased current and pyro-current measurements is a promising method to study the intrinsic ferroelectric properties in multiferroic materials.
\end{abstract}

%\begin{keyword}
%\texttt{A. Oxide; D. Multiferroic; D. Ferroelectricity; E. Pyroelectric current}
%\end{keyword}

%\vspace{2pc}
%\noindent{\it Keywords: Multiferroic, ferroelectricity, oxide, pyroelectric current}

\maketitle
\newpage

\section{Introduction} Since the discovery of spin-induced ferroelectricity in $R$MnO$_3$ ($R$ = Gd, Tb and Dy), there has been an immense effort to find out new single phase magnetoelectric multiferroic materials in which a coupled ferroelectricity and magnetism coexist~\cite{kimura, katsura, staruch, kimura1, cheong, tokunaga, kimura2}. These materials have received much attention because of their interesting fundamental science and their potential for device applications~\cite{schmid,kleemann,bea,spaldin,wang,dho}. But finding new multiferroic materials with an efficient magnetoelectric coupling at room temperature is  a challenging task because of difficulties in combining ferroelectricity and magnetism in the same phase~\cite{hill}. So far, there are only a handful of families of materials have been found to show this phenomenon in which either the ferroelectric transition is far below room temperature or the polarization is lower than that required for applications~\cite{tkimura,cheong}. Besides, the inherent leakage current in most of these materials becomes a major issue for studying their ferroelectric properties~\cite{kitagawa,feng}.

Usually, the standard Sawyer-Tower circuit has been used for the measurement of $P-E$ loop to determine ferroelectric properties in conventional ferroelectric materials which are good insulators~\cite{sawyer,scott}. However, this method seems to be not suitable for multiferroic materials because of their  leakage current and  feeble ferroelectric polarization. Recently, a modified $P-E$ loop method known as Positive-Up and Negative-Down (PUND) has been employed in determining the intrinsic ferroelectric polarization by separating out the leakage contribution~\cite{fukunaga,feng,chai}. However, the most widely used method to study ferroelectric properties in multiferroic materials is the measurement of pyroelectric current~\cite{kimura,lawes,heyer,jodlauk,johnson}. Though the measurement of pyro-current is simple, the analysis becomes complicated when the sample contains thermally stimulated free charge (TSFC) carriers.

Here, we report on the formation of an internal electric field ($E_{int}$) by TSFC carriers and its effect on the pyro-current peak at $T_{C}$ in a polycrystalline sample of the well-known multiferroic, TbMnO$_3$. In this compound, the Mn$^{3+}$ moments undergo an incommensurate sinusoidal antiferromagnetic ordering at 42 K and a commensurate cycloidal ordering at 28 K below which a spontaneous electric polarization appears because of breaking of inversion symmetry due to inverse Dzyaloshinskii- Moriya interaction~\cite{kimura,katsura,staruch,kimura2}. When the sample is poled across the ferroelectric transition in a wide temperature range ($T_{Pole}$: 130 - 10 K) by a negative external electric field ($E_{ext}$ = $-$ 4 kV/cm), we observe two pyro-current peaks in the positive direction ($+$ $I_{Pyro}$). The one at the ferroelectric transition ($T_C$ $\sim$ 26 K) is sharp whereas the other peak located in paraelectric state ($\sim$ 80 K) is relatively broad and symmetric. This broad peak is a manifestation of the existence of $E_{int}$ field. Intriguingly, the sample poled only in the paraelectric state ($T_{Pole}$: 130 - 50 K) by the same negative field ($-$ $E_{ext}$) also exhibits a pyro-current peak at $T_C$ with the same polarity ($-$ $I_{pyro}$). These unusual behaviors of pyro-current is explained by a positive internal field ($+$ $E_{int}$) generated during the poling process.  Further, it has been shown that DC-biased current measurements are useful in understanding the behavior of pyro-current and thus identifying true ferroelectricity.
\begin{figure}[t!]
\centering
\includegraphics[scale=0.4]{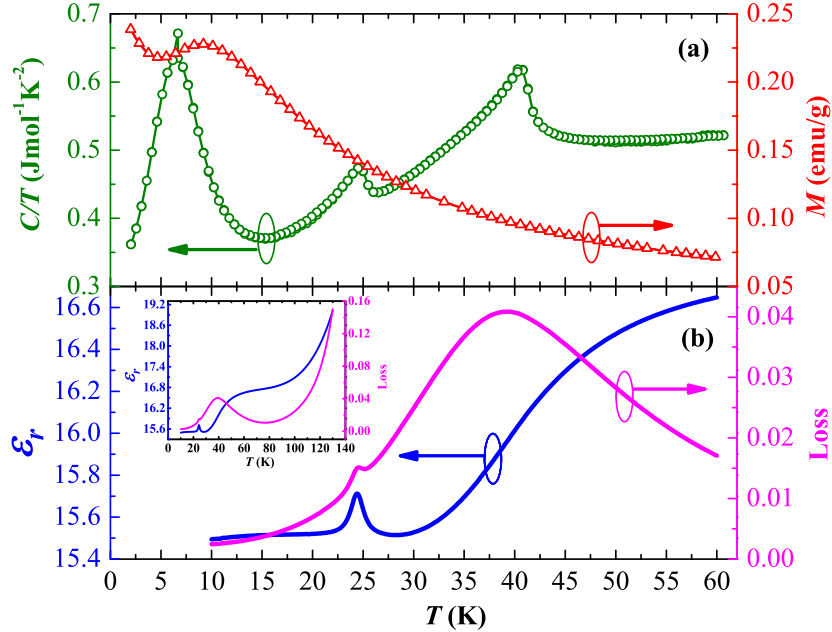}
\caption{\label{fig:sine}(a) Specific heat divided by temperature ($C/T$) (left axis) and magnetization ($M$)  (right axis) versus temperature ($T$). (b) Dielectric constant ($\epsilon_r$) (left axis) and loss ($\tan\delta$) (right axis) as a function of temperature. Inset shows the dielectric data up to 130 K.}
\end{figure}

\section{Experiments} Polycrystalline sample of TbMnO$_3$ was prepared by the standard solid state synthesis route. Phase purity was confirmed by analyzing the powder  X-ray  diffraction  data collected  using  Bruker  D8 Advance - X-ray  diffractometer.  Magnetic measurements were carried out on a SQUID VSM magnetometer and heat capacity and electrical measurements with a Physical Property Measuring System (PPMS), Quantum Design, USA. Pyroelectric and DC-biased current measurements were made with Keithley Electrometer/High resistance meter (model 6517A). Dielectric properties were measured using Agilent (E4980A) Precision LCR meter. Electrical contacts to the sample were made using silver paint.

\section{Results and Discussion} Fig. 1(a) shows specific heat divided by temperature ({\it C/T}) and magnetization ($M$), measured under field cooled condition (100 Oe), as a function of temperature ($T$). Since our sample is polycrystalline and the Tb$^{3+}$ moments are higher than Mn$^{3+}$ moments, we do not observe magnetization anomalies associated with ordering of Mn$^{3+}$ ions ~\cite{staruch1}. However, these transitions can be seen clearly in ({\it C/T}) data, including the Tb$^{3+}$ ordering at 7 $K$. Results of dielectric measurements carried out at 50 kHz are shown in Fig. 1(b) where the dielectric constant and loss data show  anomaly at $T_C$. The inset of this figure shows the dielectric data up to 130 K where the two step-like feature above ~ 30 K can be explained by small polaron tunneling and Maxwell-Wagner-type dielectric relaxation ~\cite{silveira,park}.

\begin{figure}[t!]
\centering
\includegraphics[scale=0.4]{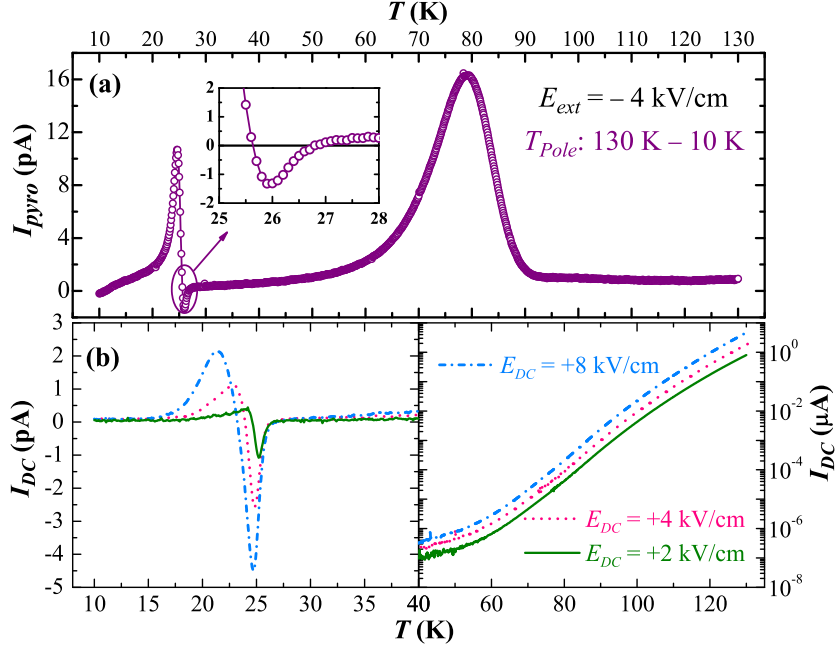}
\caption{\label{fig:sine} (a) Pyro-current versus temperature data after poling at $E_{ext}$ = $-$ 4 kV/cm and $T_{Pole}$: 130 - 10 K. (b) Variation of  DC-biased current ($I_{DC}$) with temperature at different bias field in two different temperature range with the left axis on linear scale and the right axis on a log scale.}
\end{figure}

Fig. 2(a) shows pyro-current ($I_{Pyro}$) data recorded while warming the sample at 4 K/min from 10 K to 130 K, after poling the sample from 130 K to 10 K (T$_{Pole}$) with $E_{ext}$ = $-$ 4 kV/cm. Before recording the pyro-current, electrode wires were shortened for 30 min at 10 K to remove stray charges at the electrodes. We observe two distinct positive pyro-current peaks; the low temperature one at 26 K is sharp whereas the high temperature peak at T $\sim$ 80 K is relatively broad and symmetric. Obviously, the low temperature peak is the depolarization current due to ferroelectric transition. However, by a careful observation, we can see that the current on the high temperature side of this peak goes to negative direction, as shown in the inset, before it becomes zero at T$_C$. This behavior is anomalous because the depolarization current should remain positive for a negative poling field ($E_{ext}$ = $-$ 4 kV/cm). We suggest that the negative component of the low temperature peak is related to presence of the high temperature pyro-current peak at $\sim$ 80 K. In fact, the second peak is a manifestation of presence of a positive internal electric field ($+$ $E_{int}$) in the paraelectric state. The internal field is created by  the TSFC carriers which are polarized and frozen-in during the poling process mentioned above. During the pyro-current measurement, the depolarization of these charges exhibit a peak in pyro-current. In order to confirm the non-ferroelectric origin of this peak, we analyzed the pyro-currents measured at different heating rates. We found that the high temperature pyro-current peak shifts to high temperature with increasing heating rate ~\cite{zhang}. On the other hand, the low temperature peak remains unchanged (within instrumental resolution) which is consistent with the fact that the depolarization current due to ferroelectricity should vanish always at the ferroelectric transition temperature.

In order to further distinguish these two pyro-current peaks, we have measured the temperature dependence of DC-biased (+2, +4 and +8 kV/cm) current ($I_{DC}$) while warming the sample from 10 K and the results are shown in Fig. 2(b) in two parts.  Interestingly, at the temperature corresponding to the low temperature pyro-current peak, the $I_{DC}$ shows a broad positive and a sharp negative peak whose magnitude increases with increasing applied field. The observation of positive and negative component of this peak resembles the low temperature pyro-current peak [Fig. 2(a)], however, they differ in their mechanism of origin. While warming from 10 K, the positive peak in DC-biased current ($+$ $I_{DC}$) arises due to polarization of ferroelectric dipoles and the consecutive negative peak results from depolarization current near the ferroelectric transition. In contrast, the positive component of the low tempertaure pyro-current peak  arises from depolarization current while the negative component has its origin at the positive internal field ($+$ $E_{int}$) field. More importantly, the $I_{Dc}$ does not show any feature corresponding to the high temperature pyro-current peak in the paraelectric state confirming its non-ferroelectric origin. Thus, these results demonstrate that the DC-biased current measurement can differentiate the pyro-current peak arising from ferroelectricity and that due to TSFC carriers. In fact, we observed a broad pyro-current peak in several non-ferroelectric materials such as, Na$R$MnWO$_6$ ($R$ = La, Nd and Tb)~\cite{de}, $R$MnO$_3$ ($R$ = Nd, Sm and Eu) and a garnet Sm$_3$Fe$_5$O$_{12}$ but we did not observe corresponding peak in DC-biased current (data not shown). Therefore, the presence of pyro-current does not always indicate ferroelectricity and one must be careful while analyzing the pyro-current of leaky materials particularly when the pyro-current peak appears close to magnetic transition ~\cite{ruff,ikeda,maglione}. In fact, by analyzing the shape of pyro-current peaks observed in ferroelectric and non-ferroelectric materials, we suggest that the lambda-like pyro-current with a narrow peak width is indicative of ferroelectricity. The relatively broad and symmetric peak may originate from TSFC carriers.

Now we demonstrate that the negative component of the low temperature peak is caused by the positive internal field . For this purpose,
 we have measured pyro-current under different poling temperature range with the same negative poling field ($-$ 4 kV/cm) and heating rate adapted in Fig. 2(a) and the results are shown in Fig. 3. For the poling temperature range, $T_{Pole}$: 50 - 10 K, the pyro-current measured from 10 to 130 K is shown in Fig. 3(a). It can be seen that there is only one pyro-current peak with positive polarity ($+$ $I_{Pyro}$) due to depolarization of dipoles at the ferroelectric transition without any negative component. The fact that we do not observe the negative component in the low temperature pyro-current peak and also the absence of high temperature pyro-current peak in the paraelectric state confirm that the negative component of the first peak is related to the high temperature pyro-current peak. Further, the external poling below 50 K does not produce $E_{int}$ and therefore the observed pyro-current is the result of only the external field ($-$ $E_{ext}$). In order to study  the effect of $E_{int}$ alone, we poled the sample only in the paraelectric state ($T_{Pole}$: 130 $-$ 50 K) and shortened the electrodes at 50 K. Then the sample was cooled to 10 K in short-circuit condition across the ferroelectric transition. Since the sample was cooled in the absence of $E_{ext}$ across the ferroelectric transition, we do not expect  pyro-peak at $T_C$. In contrast, we observe a pyro-peak at $T_C$ with the same polarity ($-$ $I_{Pyro}$) as that of the ($-$) $E_{ext}$ [Fig. 3(b)]. On other hand, the  broad positive pyro-current peak in the paraelectric state (80 K) is a manifestation of presence of $E_{int}$ due to TSFC carriers as discussed before. Thus, it is obvious from these measurements that the negative pyro-current peak ($-$ $I_{Pyro}$) arises from the positive internal field ($+$ $E_{int}$) which in turn is created by negative external field ($-$ $E_{ext}$).

\begin{figure}[t!]
\centering
\includegraphics[scale=0.4]{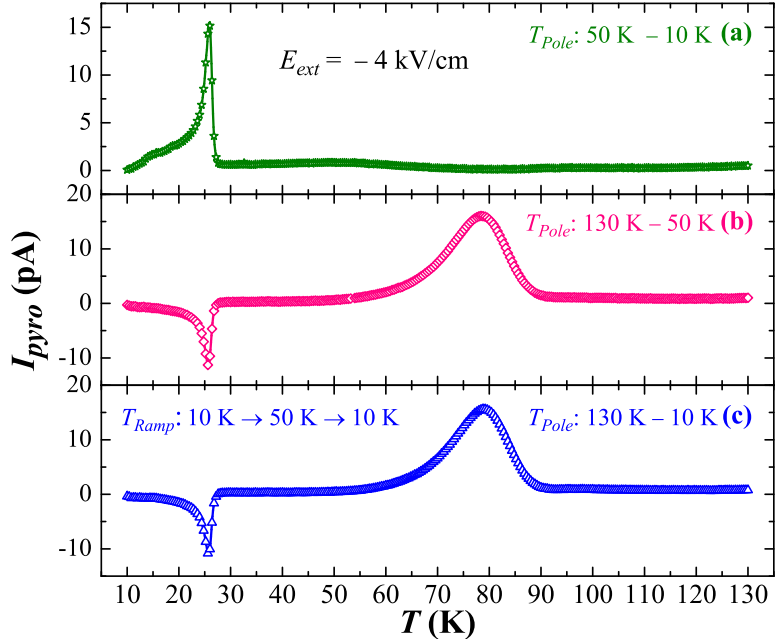}
\caption{\label{fig:sine}Temperature profile of $I_{Pyro}$ recorded under the same poling field ($E_{ext}$ = $-$ 4 kV/cm) but different poling temperature range ($T_{Pole}$). (a) $T_{Pole}$:  50 - 10 K, (b) $T_{Pole}$: 130 - 50 K and (c) $T_{Pole}$: 130 - 10 K. In last case, the sample was warmed across the $T_C$ to $T$ = 50 K in short circuit condition and cooled to 10 K prior to the $I_{Pyro}$ measurement from 10 to 130 K.}
\end{figure}

The presence of $E_{int}$ and its polarity with respect to $E_{ext}$ is further confirmed by the following warming and cooling experiments in the temperature range 10 - 50 K. After poling from 130 K to 10 K, the sample was short-circuited and warmed to a temperature (50 K) above $T_C$ and then cooled again to 10 K. In this process, the positive depolarization current ($+$ $I_{Pyro}$) would disappears at $T_C$ during the first warming cycle and therefore we should not see depolarization current in the subsequent warming measurement. Indeed, we still observe a  pyro-current peak at $T_C$ [Fig. 3(c)]with negative polarity ($-$ $I_{Pyro}$) which remains unchanged for further cooling and warming cycles between 50 K and 10 K. This peak disappears only when the sample is warmed above the high temperature peak where the $E_{int}$ field disappears. These results further confirm the presence of the $E_{int}$ field and its opposite polarity with respect to the $E_{ext}$, when the sample is poled in the paraelectric state (130 K) and cooled to 50 K or below. We have estimated the strength of $E_{int}$ with respect to the $E_{ext}$ from polarization data [Fig. 4(a)] obtained from the pyro-current data shown in Fig. 2(a) and Fig. 3. It is found that the strength of the internal poling field is three fourth of the external field ($E_{int}$ $\sim$ 0.75E$_{ext}$).

\begin{figure}[h!]
\centering
\includegraphics[scale=0.4]{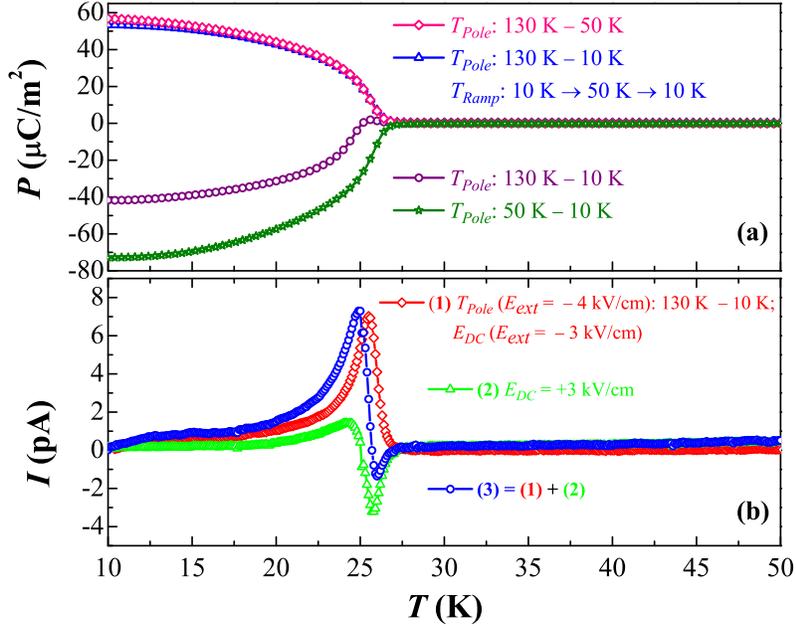}
\caption{\label{fig:sine}(a) Polarization data ($T$ $<$ 50 K) obtained from integrating the pyro-current data (Fig. 2(a) and Fig. 3) with time. Different color/symbol show the polarization for different poling procedure mentioned in the figure. (b) Results of DC-biased current ($I_{DC}$) measurements under different bias field ($E_{DC}$): (1). $I_{DC}$ ({\color{red}$\diamond$}) measured under $E_{DC}$ = ($-$ 3 kV/cm) after poling from 130 to 10 K with $-$ 4 kV/cm. (2). $I_{DC}$ ({\color{green}$\triangle$}) measured under $E_{DC}$ = $+$ 3 kV/cm. (3). Data ({\color{blue}$\circ$}) obtained by adding the $I_{DC}$ data 1 ({\color{red}$\diamond$}) and 2 ({\color{green}$\triangle$}) which is equivalent to $I_{Pyro}$ in Fig. 2(a).}
\end{figure}

The behavior of pyro-current due to the $E_{ext}$ [Fig. 3(a)], $E_{int}$  [Fig. 3(b)] and their combined effect [Fig. 2(a)] can be better understood from the orientation of ferroelectric domains with respect to the poling field as shown in the diagram [Fig. 5]. In this diagram, as the crystallographic directions are randomly oriented to the applied electric field, we have considered the dipolar orientation of each grain along the applied field direction. The diagram in Fig. 5(a) represents the effect of $-$ $E_{ext}$ on the orientation of ferroelectric dipoles (green color). This corresponds to the positive depolarization current as shown in Fig. 3(a). Fig. 5(b) depicts the formation of the $E_{int}$ due to TSFC carriers when the sample is cooled under the external poling field across the temperature range, 130 - 50 K. Under an applied field ($-$ $E_{ext}$), the charge carriers accumulate near the grain boundaries and frozen-in while cooling the sample and thus form electric dipoles which act as an internal poling field, below a freezing temperature ($T$ $<$ 50 K), which remain even after removing the $E_{ext}$ field. During this process, the free charge carriers move to oppositely charged electrodes, the polarity of ($\pm$) the $E_{int}$ is opposite ($\mp$) to that of the $E_{ext}$.  The effect of positive internal field ($+$ $E_{int}$) alone on the ferroelectric dipoles are shown in Fig. 5(c) in red color, which corresponds to the negative pyro-current ($-$ $I_{Pyro}$) peak observed in Fig. 3(b).

\begin{figure}[h!]
\centering
\includegraphics[scale=0.5]{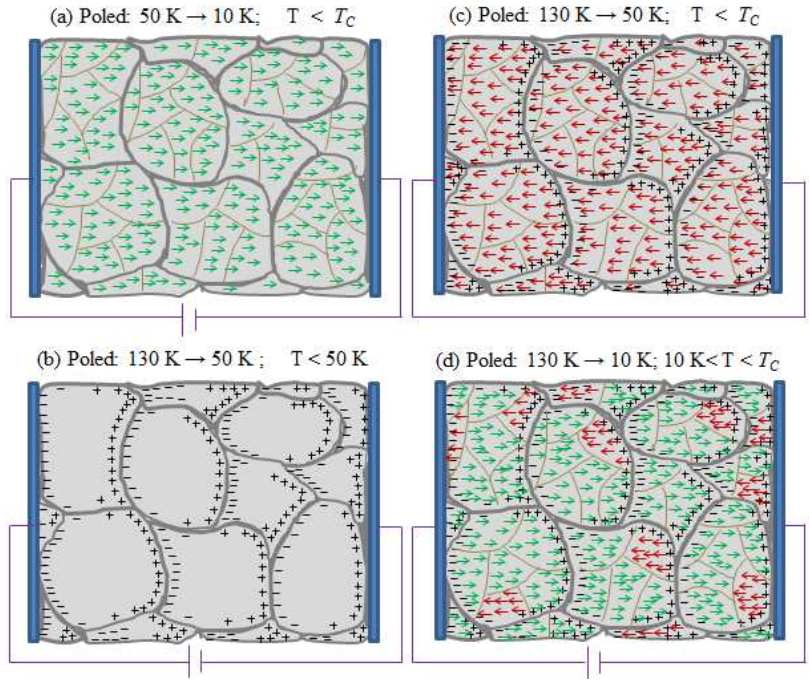}
\caption{\label{fig:sine}Schematic diagrams showing orientation of ferroelectric domains and polarization of TSFC carriers under different poling temperature range. (a) Ferroelectric dipole orientation due to external field (poled below 50 K), (b)  formation of internal field after poling the sample from 130 to 50 K, (c) ferroelectric dipole orientation due to internal electric field (poled from 130 to 50 K) and (d) oppositely aligned ferroelectric domains due to simultaneous presence of internal and external fields (Poled from 130 K to 10 K).}
\end{figure}
Finally, we discuss the behavior of pyro-current shown in Fig. 2(a), where the sample was cooled across the ferroelectric transition under both $-E_{ext}$ and $+E_{int}$ fields simultaneously. In this case, we may anticipate that the resultant pyro-current may be the effect of the sum of the pyro-current behavior under external  and internal  poling fields [Fig. 3(a) and 3(b)]. In fact, it is rather a combination of pyro-current due to the effective external field ($- E_{eff}$ = $-E_{ext}$ $+$ $E_{int}$) and DC-biased current corresponding the internal field. Fig. 5(d) represents ferroelectric domain orientation in the temperature interval 10 K $<$ T $<$ $T_C$  after the simultaneous electric poling ($-$ $E_{ext}$ and $+$ $E_{int}$), where we see that there are two oppositely aligned ferroelectric domains.  The bigger domains (green color) are formed because of the effective external field ($-E_{eff}$) during the poling (130 - 10 K) process. Upon warming in short-circuit condition, the polarized dipoles in the bigger domains would give a positive depolarization current resulting in random orientation of dipoles. At the same time, the presence of ($+E_{int}$) field will align the random oriented dipoles in the opposite direction. This situation is similar to switching of ferroelectric polarization under opposite electric field except in the present case it is driven by thermal energy.    As a result, the polarization current due to internal field will initially add up to the depolarization current. Close to the ferroelectric transition, both the depolarization and polarization current, due to switching of polarization, will cease and depolarization current resulting from  $+$ $E_{int}$ would give rise to negative component of the pyro-current peak ($-$ $I_{Pyro}$).  In order to prove this hypothesis, we have carried out two differnt DC-biased current measurements. In the first measurement, the  current data was recorded under the DC-bias field of $-$3 kV/cm after the sample was poled from 130 K to 10 K under the poling field of $-$4 kV/cm. This experiment was carried out to nullify the effect of ($+E_{int}$) by applying a negative DC-bias of magnitude equivalent to $E_{int}$ $\sim$ 0.75E$_{ext}$. This should be equivalent to pyro-current behavior under the poling field of $E_{eff}$ $\sim$ ($-$ 1 kV/cm). In the second experiment, to obtain the effect of internal field ($+E_{int}$) alone on pyro-current, the current was measured under a DC-bias of +3 kV/cm which is equivalent to $+$ $E_{int}$.   This would give rise to a polarization and depolarization current while warming the sample during the pyro-current measurement. Results of these two experiments are shown in Fig. 4(b) where we have shown the sum of these two  DC-biased current data (open circle) which reproduces the pyro-current behavior shown in Fig.2(a).

\section{Conclusion} In conclusion, pyroelectric current measurements on a polycrystalline TbMnO$_3$ demonstrate that the internal electric field generated by thermally stimulated free charge carriers manifests itself as a pyro-current peak in the paraelectric state. The differences between the nature of pyro-current due to ferroelectricity and the free charge carriers have been discussed, which would be helpful in characterizing multiferroic materials. The effect of interplay of internal and external poling fields on ferroelectric polarization has been explained by combined pyro-current and DC-biased current measurements.   More importantly, we have shown that a simple DC-biased current measurement can distinguish pyro-current peaks originating from ferroelectric polarization and free charge carriers.

\section{acknowledgement}
The authors thank the Sheikh Saqr Laboratory at the Jawaharlal Nehru Centre for Advanced Scientific Research for experimental facilities.
%\newpage

\end{document}